\documentclass[a4paper, 11pt]{article}

\usepackage[utf8]{inputenc}
\usepackage[left=2cm, right=2cm, top=2.cm, bottom=2.cm]{geometry}
\usepackage{slashed}
\usepackage{bm}
\usepackage{latexsym,amssymb,amsmath,float,url,mathrsfs,physics}
\usepackage{latexsym}
\usepackage{graphicx}
\usepackage{amsmath}
\usepackage{comment}
\usepackage{subfigure}
\usepackage{pifont}
\usepackage{blindtext}
\usepackage{adjustbox}
\usepackage{multirow}
\usepackage{tabularx}
\usepackage[table,dvipsnames]{xcolor}
\usepackage{orcidlink}
\usepackage{tikz}
\usetikzlibrary{tikzmark}
\usepackage{fontawesome}
\usepackage{booktabs}
\usepackage{blkarray}
\usepackage{graphicx}
\usepackage{amsfonts}
\usepackage{soul}
\usepackage{amssymb}
\usepackage{cancel}
\usepackage{tcolorbox}
\usepackage{tensor}
\usepackage{textcomp}
\usepackage{graphics,appendix,afterpage,makecell} 

\usepackage[
  backend=biber, style=numeric-comp, giveninits=true, doi=false,
  sorting=none, backref,backrefstyle=none, maxnames=7, minnames=6
]{biblatex}
\AtBeginBibliography{\small}  
  
\DeclareFieldFormat[article]{title}{\textit{#1}} 
\DeclareFieldFormat[incollection]{title}{\textit{#1}} 
\DeclareFieldFormat[conference]{title}{\textit{#1}} 
\DeclareFieldFormat[inproceedings]{title}{\textit{#1}} 
\DefineBibliographyStrings{english}{ 
  backrefpage = {p\adddot\addspace}, 
  backrefpages = {pp\adddot\addspace} }
 
\renewbibmacro{in:}{%
  \ifentrytype{incollection}{%
   \printtext{\bibstring{in}\intitlepunct}%
   }{\addcomma\space }%
   }
\renewbibmacro*{doi+eprint+url}{
  \setunit{\addcomma\addspace}\newblock 
  \iftoggle{bbx:doi}
    {\printfield{doi}}
    {\printfield{doi}}%
  \setunit{\addcomma\addspace}\newblock 
  \iftoggle{bbx:eprint}
    {\usebibmacro{eprint}}
    {}%
  \setunit{\addcomma\addspace}\newblock
  \iftoggle{bbx:url}
    {\usebibmacro{url+urldate}}
    {}}
\DeclareFieldFormat{doi}{%
  \ifhyperref
    {\href{https://doi.org/#1}{\textsc{doi}}}
    {doi}}
\makeatletter
\DeclareFieldFormat{eprint:arxiv}{%
  arXiv\addcolon\space
  \ifhyperref
    {\href{http://arxiv.org/\abx@arxivpath/#1}{%
       #1
       \iffieldundef{eprintclass}
     {}
     {\addspace\mkbibbrackets{\thefield{eprintclass}}}}}
    {#1
     \iffieldundef{eprintclass}
       {}
       {\addspace\mkbibbrackets{\thefield{eprintclass}}}}}
\makeatother
\renewbibmacro*{publisher+location+date}{%
  \printlist{publisher}%
  \iflistundef{location}
    {\setunit*{\addcomma\space}}
    {\setunit*{\space(}}%
  \printlist{location}%
  \setunit*{)\addcomma\space}%
  \usebibmacro{date}%
  \newunit}
\DeclareSourcemap{
 \maps[datatype=bibtex,overwrite=true]{
  \map{
    \step[fieldsource=Collaboration, final=true]
    \step[fieldset=usera, origfieldval, final=true]
  }}}
\renewbibmacro*{author}{%
  \iffieldundef{usera}{%
  \printnames{author}%
  }{%
  \textbf{\printfield{usera}} collaboration, \printnames{author}%
  }}
  
\usepackage{url}
\definecolor{TardisBlueAcceso}{RGB}{0, 80, 200}

\definecolor{oucrimsonred}{rgb}{0.6, 0.0, 0.0}
\definecolor{persianblue}{rgb}{0.11, 0.22, 0.73}
\definecolor{forestgreen}{rgb}{0.13,0.35,0.13}
\definecolor{lightgray}{rgb}{0.83, 0.83, 0.83}
\definecolor{verdes}{rgb}{0.1, 0.5, 0.1}%
\definecolor{cornellred}{rgb}{0.7, 0.11, 0.11}

\definecolor{VioletRed4}{rgb}{0.55, 0.13, .32}
\definecolor{rossocorsa}{rgb}{0.83, 0.0, 0.0}

\usepackage{hyperref}
\hypersetup{colorlinks, citecolor=ForestGreen, linkcolor=BrickRed, urlcolor=NavyBlue}
\usepackage[capitalise]{cleveref}

\begin{document}

\title{\bf Exact solution of the Einstein-scalar-Gauss-Bonnet model with Noether symmetry constraints}

\author{
Olga Razina$^1$\footnote{olvikraz@mail.ru} \orcidlink{0000-0002-4400-4789},
Dauren Rakhatov$^{1}$\footnote{godauren@gmail.com (corresponding author)} \orcidlink{0000-0002-5318-2990}, 
Pyotr Tsyba$^1$\footnote{pyotrtsyba@gmail.com} \orcidlink{0000-0003-4928-0392} and 
Emilio Elizalde$^2$\footnote{elizalde@ice.csic.es} \orcidlink{0000-0003-2556-8930}}

\date{}
\maketitle

\begin{center}
{\small
$^1$~L.N.Gumilyov Eurasian National University, Astana, 010008, Kazakhstan\\
$^2$~Institute of Space Sciences, ICE/CSIC and IEEC, Campus UAB, Bellaterra, 08193, Spain}
\end{center}

\begin{abstract} 
\noindent
By applying Noether symmetry methods, analytic solutions are obtained for a generalized Einstein-scalar-Gauss-Bonnet model with a $\xi(\phi)f(G)$ component. Variation with respect to the metric, supplemented by small perturbations, produces the equations of motion and the terms that determine the propagation speed of tensor perturbations. The resulting Hubble parameter incorporates contributions from stiff matter and dark energy, the last originating from a scalar field non-minimally coupled to the Gauss-Bonnet invariant. The viability of the model is assessed by using Cosmic Chronometers, Baryon Acoustic Oscillations, and type Ia supernovae data. Best model selection based on information criteria indicates a slight preference for this new framework over the $\Lambda$ Cold Dark Matter model. Stability of the model follows from the positive speed of sound and absence of ``Ostrogradsky ghosts''. The total equation of state parameter hints towards the presence of a transition from decelerated to accelerated expansion at $z\approx 0.66$, corresponding to the transition from matter to dark energy dominance. Early Universe dynamics, derived from the slow-roll parameters, spectral indices, and the tensor-to-scalar ratio, are found to be perfectly consistent with observations from Planck 2018 and the Atacama Cosmology Telescope.
\end{abstract}

\section{Introduction}
Cosmological observations clearly indicate the accelerated expansion of the Universe \cite{Riess1998ObservationalConstant, Perlmutter1998MeasurementsSupernovae, Scolnic2022TheRelease}. Within the framework of the standard $\Lambda$ Cold Dark Matter ($\Lambda$CDM) model, such accelerated expansion is explained by the presence of an hypothetical dark energy, which accounts for approximately 68\% of the energy content of the Universe. However, the nature and the origin of dark energy is still unknown \cite{DiValentino2025ThePhysics}. Together with dark matter, which makes up some 27\%, the invisible components account for about 95\% of the contents of the Universe \cite{Farnes2018AFramework, Hinshaw2013Nine-YearResults,Bamba2012DarkTests}. Clarifying the nature of the invisible part of the Universe and  its impact on the dynamics of expansion are most urgent tasks to do. \cite{Chaudhary2026IsCrisis,Perivolaropoulos2022ChallengesUpdate, Odintsov2025ModifiedOver}. Aiming to solve these problems, active search is underway for alternatives and modifications of the standard cosmological model, which could explain the dynamics of the Universe expansion without the need for exotic components \cite{Nojiri2011UnifiedModels, Nojiri2017ModifiedEvolution, Shankaranarayanan2022ModifiedWhat}. An additional difficulty that arises when modifying theories of gravity is the construction of universal cases, capable of simultaneously describing both the early Universe and its later accelerated expansion \cite{Verde2019TensionsUniverse}.

Scalar-tensor models occupy a special place among modified gravity theories available \cite{Fujii2003TheGravitation}. Gravitational phenomena are there described not only by geometric features of space, but also by the introduction of additional scalar fields. The combined contribution of tensor and scalar components make it possible to build a universal model without the need to introduce dark energy and fine-tune the initial conditions. 

The Einstein-scalar-Gauss-Bonnet model (EsGB) is among the most studied scalar-tensor theories \cite{Nojiri2005Gauss-BonnetEnergy, Oikonomou2024EinsteinGaussBonnetEra, Tsyba2025TheModel, Fernandes2022TheReview}. Its action, in natural units $8\pi G=M_p^{-2}=\hbar=c=1$, reads
\begin{equation}\label{standardEsGB}
	S = \int d^4x \sqrt{-g} \left[ \frac{R}{2 \kappa^2} - 
    \frac{1}{2} \partial_\mu\phi \partial^\mu\phi - V(\phi) - 
    \xi(\phi)G\right], 
\end{equation}
where $g$ is the determinant of the metric tensor, $R$ the scalar space curvature, $\phi$ a scalar field, $V(\phi)$ the potential of the scalar field, $\xi(\phi)$ the coupling function, and $G$ is the Gauss-Bonnet (GB) invariant.

In this type of models, a non-minimal coupling with a scalar field allows for adjusting the contribution of the GB invariant. At the same time, actions of this type have been successfully applied to obtain solutions for the early and late Universe \cite{RCPH20259211, Nojiri2023UnifyingGravity, Oikonomou2025RescaledInflation}. Moreover, switching to special metrics also allows to get solutions for black holes and massive objects \cite{Q.2025CompactConfigurations, Saavedra2025NeutronGravity, Corman2024BlackGravity}.

Despite their theoretical beauty, EsGB models face severe limitations. The presence of higher-order invariants often leads to the appearance of ``ghost'' degrees of freedom, in the form of derivatives above the second order. Degrees of freedom of this kind are called ``Ostrogradsky's ghosts'', and they can lead to instabilities \cite{Klein2016ExorcisingSystems, Aoki2020GhostTheorem}. Additional components can also cause the speed of gravitational waves $c_{GW}$ to deviate from the speed of light, thus contradicting the results of the GW170817 event, which set the very tight limit $\mid\frac{c_{GW}}{c}-1\mid<10^{-15}$ \cite{Abbott2017GW170817:Inspiral, Abbott2017Gravitational170817A, Odintsov2025GW170817Data}. Such constraints narrow the space of possible model solutions and require careful consideration.

Anyway, since most of the constraints can be violated due to the non-minimal coupling of the GB invariant, a generalized model has been developed in which the standard component $\xi(\phi)G$ is replaced by $\xi(\phi)f(G)$, where $f(G)$ is an arbitrary function of the GB invariant. This modification allows to expand the set of viable solutions while preserving the ability to investigate non-trivial, physically meaningful cases. The model also incorporates a matter Lagrangian, enabling the consideration of the matter contribution and its impact on the Universe expansion.

The aim of this work is to derive an analytic solution that aligns with observational data and the theoretical constraints. To this end, it is most important to identify physically viable cases of the functions $\xi(\phi)$ and $f(G)$. Generically, the studies of the EsGB model are based on the reconstruction method, which involves an a priori choice of functions \cite{Hussain2024InteractingGravity, Biswas2024Gauss-BonnetWaves, Millano2023GlobalMatter}. This approach makes it easier to  obtain solutions of the equations of motion and often avoids  contradiction with the observational data. However, this approach is not suited for deriving functional dependencies directly from the action, which leads to a considerable increase in the number of assumptions and to a decrease in the reliability of the obtained solutions. 

Unlike those reconstructive approaches, analytic solutions do allow to obtain functional dependencies directly, without introducing hypothetical forms for the functions. This approach ensures self-consistency, and makes it possible to identify non-obvious internal patterns of the model, which could otherwise have been missed  during the reconstruction process. However, obtaining analytic solutions is rather difficult, owing to the non-linear form of the system of equations of motion.

Obtaining analytic solutions can be simplified by calculating additional model constraints. Having more equations allows for the calculation of the additional functional dependencies caused by physical limitations \cite{Dialektopoulos2018NoetherGravity, Capozziello2014NoetherCosmology, Capozziello2012HamiltonianCosmology}. To this end, in the present work methods based on Noether symmetry will be used. This approach makes it possible to obtain a conserved quantity, which can then be used as an additional constraint, thereby making it easier to obtain analytic solutions. In addition to the conservation law, Noether symmetry methods make it possible to calculate unknown functions \cite{Bajardi2020FGCosmology, Fazlollahi2018FRModel}. 

As a result, it will be possible to obtain the values of the non-minimal coupling function and of the GB invariant. We will prove that the conserved quantity is actually constant in time, and that it is consistent with the equations of motion of the model. Using the  values thus obtained in the Lagrangian will lead to the disappearance of ``ghosts'', as time derivatives above second order are absent. The final solution has the form of the Hubble parameter depending on the redshift \cite{Klein2016ExorcisingSystems, Aoki2020GhostTheorem}. An important feature of the solutions obtained is a certain degree of similarity with the standard $\Lambda$CDM model. However, the dark energy component in the  solutions here  appears naturally, from the equations of motion, while being two-component ones. Additionally, the model contains a stiff matter component $\Omega_s(1+z)^6$ \cite{Chavanis2014CosmologyEra, Lankinen2016GravitationalUniverse} and a coupling constant $\phi_0$, whose main contribution is concentrated in the early Universe. 

This work is divided into five sections. After this introductory one, in Section \ref{sec2}, the action of the model and the equations of motion are introduced, and the condition for limiting the speed of gravitational waves is calculated. Section \ref{sec3} is devoted to the calculation of analytic solutions, using Noether symmetry methods. In Section \ref{sec4}, the methods of model verification, by comparison with observational data, are discussed in detail. Finally, the results of the model validation and their analytic interpretation are presented in Section \ref{sec5}.

\section{Field Equations and the Propagation of Gravitational Waves}\label{sec2}

The model under consideration is based on the standard action of the EsGB theory (\ref{standardEsGB}), but the non-minimal coupling of the scalar field with the GB invariant is generalized to the form $\xi(\phi)f(G)$.
\begin{equation}\label{actionmodel}
    S = \int d^4x \sqrt{-g} \left[ \frac{R}{2 \kappa^2} - 
    \frac{1}{2} \partial_\mu\phi \partial^\mu\phi - V(\phi) - 
    \xi(\phi) f(G) + L_m \right], 
\end{equation}
where $f(G)$ is a function of the GB invariant and $L_m$ is the Lagrangian density of matter. The scalar curvature $R$ and the GB invariant $G$ are 
\begin{equation}\label{riccigaussbonnet}
    R = g^{\mu\nu}R_{\mu\nu},  \;\;\;\; G = R_{\mu\nu\rho\sigma}R^{\mu\nu\rho\sigma}-4R_{\mu\nu}R^{\mu\nu}+R^2,
\end{equation}
where $R_{\mu\nu}$ is the Ricci tensor and $R_{\mu\nu\rho\sigma}$ is the Riemann tensor. Varying the action (\ref{actionmodel}) with respect to the metric tensor $g_{\mu\nu}$ using the definitions (\ref{riccigaussbonnet}) leads to the field equation. In this context, the functions denoted as $V = V(\phi)$, $f = f(G)$, and $\xi = \xi(\phi)$ preserve their respective functional dependencies. However, for the sake of conciseness, these functions may be expressed without explicitly stating their arguments.
\begin{align}
0&= \frac{1}{2\kappa^2} \left( \frac{1}{2}g_{\mu\nu}R - R_{\mu\nu} \right)
+ \frac{1}{2}\partial_\mu \phi \partial_\nu \phi 
- \frac{1}{4}g_{\mu\nu} \partial_\rho \partial^\rho \phi 
- \frac{1}{2} g_{\mu\nu} V - \frac{1}{2}g_{\mu\nu} \xi \left( f - G f_G \right)
\nonumber \\
& - 2 \left( \nabla_\mu \nabla_\nu \xi f_G \right)R 
+ 2g_{\mu\nu} \left( \nabla^2 \xi f_G \right)R + 4 \left( \nabla_\rho \nabla_\mu \xi f_G \right)R^\rho_\nu + 4 \left( \nabla_\rho \nabla_\nu \xi f_G \right)R^\rho_\mu
\nonumber \\
& - 4 \left( \nabla^2 \xi f_G \right)R_{\mu\nu} - 4g_{\mu\nu} \left( \nabla_\rho \nabla_\sigma \xi f_G \right)R^{\rho\sigma} 
+ 4 \left( \nabla^\rho \nabla^\sigma \xi f_G \right)R_{\mu\rho\nu\sigma} 
+ \frac{1}{2}T^{(m)}_{\mu\nu}, \label{fieldeq}
\end{align}
where $f_G = \frac{d}{dG}f(G)$ and $\nabla^2 = \nabla^\mu \nabla_\mu = g^{\mu\nu}\nabla_\nu \nabla_\mu$. A similar variation, but with respect to a scalar field $\phi$, results in modified Klein-Gordon equation
\begin{equation}\label{SFfieldeq}
    0 = \nabla^2\phi - V_\phi - \xi_\phi f.
\end{equation}
Since the EsGB models can lead to deviations of the gravitational wave speed from the speed of light, the next step is to derive the field equation considering small perturbations $h_{\mu\nu}\ll1$ \cite{Nojiri2024PropagationSpacetimes, Kausar2016GravitationalTheories}. This approach allows for the analysis of the propagation speed of tensor perturbations and verifies compliance with observational constraints on the gravitational wave speed. For further variation, the following transition is applied:
\begin{equation}
    g_{\mu\nu} \rightarrow g_{\mu\nu} + h_{\mu\nu}.
\end{equation}

This work considers transversely propagating gravitational waves without additional degrees of freedom, which is expressed as a transverse-traceless gauge \cite{Nojiri2024PropagationGravity, Butcher2012LocalizingGravity}
\begin{equation}
    \nabla^\mu h_{\mu\nu} = 0, \;\;\; g^{\mu\nu} h_{\mu\nu} = 0.
\end{equation}

The condition $\mathrm{\nabla}^\mu h_{\mu\nu}=0$ eliminates non-physical gauge degrees of freedom. Combined with the traceless condition $g^{\mu\nu}h_{\mu\nu}=0$, it helps isolate the transverse components of the gravitational wave. The traceless condition removes the scalar polarization mode. As a result, only two physical degrees of freedom remain, corresponding to the massless spin-two mode \cite{Nojiri2024PropagationGravity}. A variation of Eq.~(\ref{fieldeq}) in accordance with these constraints yields

\begin{align}
    0 =\ &
    \left[\frac{1}{4\kappa^2}R - \frac{1}{2}\partial_\rho\phi\partial_\rho\phi 
    - \frac{1}{2}V - 2 (\nabla^2 \xi f_G) R 
    - 4(\nabla_\rho \nabla_\sigma \xi f_G) R^{\rho\sigma} \right. 
    \nonumber\\
    &\left. - \frac{1}{2} \xi (f - f_G G + f_{GG}G^2) \right] h_{\mu\nu} 
    \nonumber\\
    &+ \left[ \frac{1}{4}g_{\mu\nu} \partial^\tau \phi \partial^\eta \phi 
    - 2g_{\mu\nu} (\nabla^\tau \nabla^\eta \xi f_G)R 
    - 4 (\nabla^\tau \nabla_\mu \xi f_G)R^\eta_\nu \right. 
    \nonumber\\
    & - 4 (\nabla^\tau \nabla_\nu \xi f_G)R^\eta_\mu 
    + 4 (\nabla^\tau \nabla^\eta \xi f_G)R_{\mu\nu} 
    + 4g_{\mu\nu} (\nabla^\tau \nabla_\sigma \xi f_G)R^{\eta\sigma} 
    \nonumber\\
    &\left. + 4g_{\mu\nu} (\nabla_\rho \nabla^\tau \xi f_G)R^{\rho\eta}
    - 4 (\nabla^\tau \nabla^\sigma \xi f_G)R^\eta_{\mu\nu\sigma} 
    - 4 (\nabla^\rho \nabla^\tau \xi f_G)R^\eta_{\mu\rho\nu} \right] h_{\tau\eta} 
    \nonumber\\
    &+ \frac{1}{2} \Big[
    2 \delta^\eta_\mu \delta^\zeta_\nu (\nabla_\kappa \xi f_G)R
    - 4 \delta^\eta_\rho \delta^\zeta_\mu (\nabla_\kappa \xi f_G)R^\rho_\nu 
    - 4 \delta^\eta_\rho \delta^\zeta_\nu (\nabla_\kappa \xi f_G)R^\rho_\mu 
    \nonumber\\
    &+ 4 g_{\mu\nu} \delta^\eta_\rho \delta^\zeta_\sigma (\nabla_\kappa \xi f_G) R^{\rho\sigma}
    - 4 g^{\rho\eta} g^{\sigma\zeta} (\nabla_\kappa \xi f_G)R_{\mu\rho\nu\sigma} \Big] 
    \nonumber\\
    &\times g^{\kappa\lambda} (\nabla_\eta h_{\zeta\lambda} + \nabla_\zeta h_{\eta\lambda} 
    - \nabla_\lambda h_{\eta\zeta}) 
    - 2 g_{\mu\nu} (\nabla^\sigma \nabla^\rho \xi f_{GG}) GR_{\mu\rho\nu\sigma} h^{\mu\nu}
    \nonumber\\
    &- \left[ \frac{1}{4\kappa^2} g_{\mu\nu} 
    - 2 (\nabla_\mu \nabla_\nu \xi f_G) 
    + 2g_{\mu\nu} (\nabla^2 \xi f_G) \right]R^{\mu\nu} h_{\mu\nu} \nonumber\\
    &+ \frac{1}{2} \Big[ 
    \Big( -\frac{1}{2 \kappa^2} - 4(\nabla^2 \xi f_G)\Big) \delta^\tau_\mu \delta^\eta_\nu 
    + 4 (\nabla_\rho \nabla_\mu \xi f_G)\delta^\eta_\nu g^{\rho\tau} \nonumber\\
    &+ 4 (\nabla_\rho \nabla_\nu \xi f_G) \delta^\tau_\mu g^{\rho\eta} 
    - 4 g_{\mu\nu} (\nabla^\tau \nabla^\eta \xi f_G) \Big] \nonumber\\
    &\times \left( - \nabla^2 h_{\tau\eta} 
    - 2R^{\lambda\varphi}_{\eta\tau} h_{\lambda\varphi} 
    + R^\varphi_\tau h_{\varphi\eta} + R^\varphi_\eta h_{\varphi\tau} \right) \nonumber \\
    &+ \frac{1}{2} \frac{\partial T^{(m)}_{\mu\nu}}{\partial g_{\tau\eta}} h_{\tau\eta} 
    + 2 (\nabla^\rho \nabla^\sigma \xi f_G)
    \left( \nabla_\nu \nabla_\rho h_{\sigma\mu} 
    - \nabla_\mu \nabla_\nu h_{\sigma\rho} \right. \nonumber\\
    &\left. - \nabla_\sigma \nabla_\rho h_{\mu\nu} 
    + \nabla_\sigma \nabla_\mu h_{\nu\rho}
    + h_{\mu\varphi} R^\varphi_{\rho\nu\sigma} 
    - h_{\rho\varphi} R^\varphi_{\mu\nu\sigma} \right) \nonumber\\
    &+ g_{\mu\nu} (\nabla_\mu \nabla_\nu \xi f_{GG})GR h^{\mu\nu} 
    - 4 (\nabla_\rho \nabla_\nu \xi f_{GG})GR h^\rho_\mu 
    + 2(\nabla^2 \xi f_{GG})GR h_{\mu\nu} \nonumber \\
    &- \left[ 2\nabla_\mu \nabla_\nu h^{\mu\nu}R 
    - 4R^{\mu\nu} \left( \nabla^2 h_{\mu\nu} - 2 \nabla_\rho \nabla_\nu h^\rho_\mu \right) 
    + 4 \nabla^\sigma \nabla^\rho h^{\mu\nu} R_{\mu\rho\nu\sigma} \right] \nonumber \\
    &\times \Big[-2 (\nabla_\mu \nabla_\nu \xi f_{GG})R 
    + 2g_{\mu\nu} (\nabla^2 \xi f_{GG})R 
    + 4 (\nabla_\rho \nabla_\mu \xi f_{GG})R^\rho_\nu \nonumber\\
    &+ 4 (\nabla_\rho \nabla_\nu \xi f_{GG})R^\rho_\mu 
    - 4g_{\mu\nu} (\nabla_\rho\nabla_\sigma\xi f_{GG})R^{\rho\sigma} 
    + 4 (\nabla^\rho \nabla^\sigma \xi f_{GG}) R_{\mu\rho\nu\sigma} \Big].
\label{fieldeqhmunu}
\end{align}

Observations from the multi-messenger event GW170817 constrain the deviation of the gravitational wave speed from the speed of light to $|c^2_T-c^2|<6\cdot10^{-15}$ \cite{Abbott2017GW170817:Inspiral, Abbott2017Gravitational170817A, Odintsov2025GW170817Data}. To investigate possible deviations from this bound, it is sufficient to identify in Eq.~(\ref{fieldeqhmunu}) the terms that contain second-order derivatives of the perturbation tensor \cite{Elizalde2024PropagationTheory}.
\begin{align}
    I_{\mu\nu} &= I^{(1)}_{\mu\nu} + I^{(2)}_{\mu\nu}, \nonumber\\
    I^{(1)}_{\mu\nu} &= \frac{1}{2} \Big[
    \Big( -\frac{1}{2\kappa^2} - 4\nabla^2 \xi f_G \Big)\delta^\tau_\mu \delta^\eta_\nu 
    + 4 (\nabla_\rho \nabla_\mu \xi f_G)\delta^\eta_\nu g^{\rho\tau} 
    + 4 (\nabla_\rho \nabla_\nu \xi f_G) \delta^\tau_\mu g^{\rho\eta} \nonumber\\
    &\quad - 4 g_{\mu\nu} \nabla^\tau \nabla^\eta \xi f_G 
    \Big] \nabla^2 h_{\tau\eta}, \nonumber\\
    I^{(2)}_{\mu\nu} &= 2 \nabla^\rho \nabla^\sigma \xi f_G \big(
    \nabla_\nu \nabla_\rho h_{\sigma\mu}
    - \nabla_\mu \nabla_\nu h_{\sigma\rho}
    - \nabla_\sigma \nabla_\rho h_{\mu\nu}
    + \nabla_\sigma \nabla_\mu h_{\nu\rho}
    \big).
\label{I_munu}
\end{align}

Although both components contain second-order derivatives, only $I_{\mu\nu}^{\left(2\right)}$ can lead to deviations from the speed of light. The component $I_{\mu\nu}^{\left(1\right)}$ contains the d'Alembertian operator $\nabla^2$, which acts independently on each component of $h_{\tau\eta}$ and does not introduce directional dependence. In contrast, $I_{\mu\nu}^{\left(2\right)}$ contains mixed derivatives $\nabla_\nu \nabla_\rho h_{\sigma\mu}$, which couple different components of the perturbation tensor \cite{Nojiri2024PropagationGravity, Nojiri2024PropagationSpacetimes}. The absence of deviations in the speed of gravitational waves is possible under the condition
\begin{equation}\label{propcond}
    \nabla_\mu \nabla^\nu \xi f_G = 
    \frac{1}{4}g_{\mu\nu}\nabla^2\xi f_G.
\end{equation} 

This condition ensures that all direction-dependent terms in $I^{(2)}_{\mu\nu}$ cancel out, reducing the modified wave equation to the standard form. However, in a general spacetime, the condition (\ref{propcond}) cannot be satisfied. In general relativity, it is satisfied  by a maximally symmetric spacetime, only. However, there exists a class of Friedmann-Robertson-Walker (FRW) spacetime metrics that satisfy these conditions due to their isotropy and homogeneity \cite{Dussault2020AEquations, Faraoni2011AFluids}.
\begin{equation}\label{FRW}
    ds^2 = -dt^2 + a(t)^2 \left( dx^2 + dy^2 + dz^2 \right),
\end{equation}
where $t$ denotes the cosmic time and $a(t)$ is the scale factor. Using the FRW metric and the functional dependencies $\xi(\phi(t))$ and $f(G(t))$, the condition (\ref{propcond}) can be reduced to
\begin{equation}\label{Gwconstraint}
    (\xi f_G)_{tt} - H(\xi f_G)_t = 0,
\end{equation}
where $H = \frac{\dot{a}}{a}$ is the Hubble parameter, and the dot and $(\xi f_G)_t$ denote derivatives with respect to cosmic time. Setting the metric (\ref{FRW}) as background allows deriving the equations of motion from Eqs. (\ref{fieldeq}-\ref{SFfieldeq}), where the pressure of matter vanishes $p_m=0$.
\begin{equation}\label{Friedmann1}
    3H^2 + 2\dot{H} + \frac{1}{2}\dot{\phi}^2 - V - \xi(f-f_GG) - 
    16H(\dot{H} + H^2)(\xi f_G)_t - 8H^2 (\xi f_G)_{tt} = 0,
\end{equation}
\begin{equation}\label{Friedmann2}
    - 3H^2 + \frac{1}{2}\dot{\phi}^2 + V + \xi(f-f_GG) - 
    24H^3(\xi f_G)_t + \rho_m = 0,
\end{equation}
\begin{equation}\label{Klein-Gordon}
    \ddot{\phi} + 3H\dot{\phi} + V_\phi + \xi_\phi f = 0.
\end{equation}

Observe that the system of equations (\ref{Friedmann1}-\ref{Klein-Gordon}) is highly non-linear and contains many unknown components. The presence of first- and second-order time derivatives of $\xi f_G$ introduces additional complexity due to the coupling between the gravitational and scalar parts. For this reason, instead of a direct solution, methods based on Noether symmetry will be used. This approach allows for identifying conservation laws and limiting possible solutions.

\section{Application of Noether Symmetry Methods}\label{sec3}

Noether symmetry methods provide the basis for determining conservation laws and obtaining exact solutions to the equations of motion. To apply them, it is necessary to isolate the Lagrangian from the action (\ref{actionmodel}) using a metric (\ref{FRW}).
\begin{equation}\label{Lagrangian}
    L = -3 \dot{a}^2 a +\frac{1}{2}a^3\dot{\phi}^2 - a^3V - 
    a^3\xi(f-f_GG) + 8\dot{a}^3 (\xi f_G)_{t} + a^3L_m
\end{equation}
According to Noether's theorem, the Lagrangian admits the presence of symmetry under the condition $XL= 0$ \cite{Capozziello2007ReconstructionSymmetry}, where $X$ is the Lie derivative with prolongations \cite{Bajardi2022NoetherCosmology, Capozziello2016NoetherCosmology}, which is calculated by the formula
\begin{equation}\label{LieDerivative}
    X = \beta \frac{\partial}{\partial a} + 
    \gamma\frac{\partial}{\partial G} + 
    \delta \frac{\partial}{\partial \phi} + 
    \dot{\beta} \frac{\partial}{\partial \dot{a}} + 
    \dot{\gamma} \frac{\partial}{\partial \dot{G}} + 
    \dot{\delta} \frac{\partial}{\partial \dot{\phi}},
\end{equation}
where the prolongations have the form as follows
\begin{align}
    \dot{\beta} &= \dot{a} \frac{\partial \beta}{\partial a} + 
    \dot{G} \frac{\partial \beta}{\partial G} +
    \dot{\phi} \frac{\partial \beta}{\partial \phi}, \nonumber\\
    \dot{\gamma} &= \dot{a} \frac{\partial \gamma}{\partial a} + 
    \dot{G} \frac{\partial \gamma}{\partial G} +
    \dot{\phi} \frac{\partial \gamma}{\partial \phi}, \nonumber\\
    \dot{\delta} &= \dot{a} \frac{\partial \delta}{\partial a} + 
    \dot{G} \frac{\partial \delta}{\partial G} +
    \dot{\phi} \frac{\partial \delta}{\partial \phi}.
\label{Prolongation}
\end{align}

Next, the operator (\ref{LieDerivative}) is applied to the Lagrangian (\ref{Lagrangian}) with the symmetry condition $XL=0$
\begin{align}
\nonumber
XL &= -3\dot{a}^2\beta + \tfrac{3}{2}a^2\dot{\phi}^2\beta - 3a^2V\beta 
- 3a^2\xi f\beta + 3a^2\xi Gf_G\beta + a^3\xi f_{GG}G\gamma \\
\nonumber
&\quad + 8\dot{a}^3\xi_\phi\dot{\phi}f_{GG}\gamma + 8\dot{a}^3\xi f_{GGG}\dot{G}\gamma 
- a^3V_\phi\delta - a^3\xi_\phi f\delta + a^3\xi_\phi Gf_G\delta \\
\nonumber
&\quad + 8\dot{a}^3\xi_{\phi\phi}\dot{\phi}f_G\delta 
+ 8\dot{a}^3\xi_\phi f_{GG}\dot{G}\delta - 6\dot{a}\dot{\phi}a\beta_\phi 
+ 24\dot{a}^2\xi_\phi\dot{\phi}^2 f_G \beta_\phi \\
\nonumber
&\quad + a^3\dot{\phi}^2\delta_\phi + 24\dot{a}^2\xi\dot{\phi}f_{GG}\dot{G}\beta_\phi 
- 6\dot{a}^2a\beta_a + 24\dot{a}^3\xi_\phi\dot{\phi}f_G\beta_a \\
\nonumber
&\quad + 24\dot{a}^3\xi f_{GG}\dot{G}\beta_a - 6\dot{a}\dot{G}a\beta_G 
+ 24\dot{a}^2\xi_\phi\dot{\phi}f_G\dot{G}\beta_G 
+ 24\dot{a}^2\xi f_{GG}\dot{G}^2\beta_G \\
\nonumber
&\quad + 8\dot{a}^4\xi f_{GG}\gamma_a + 8 \dot{a}^3\xi f_{GG}\dot{G}\gamma_G 
+ 8\dot{a}^3\xi\dot{\phi}f_{GG}\gamma_\phi + \dot{a}\dot{\phi}a^3\delta_a \\
&\quad + 8\dot{a}^4 \xi_\phi f_G\delta_a + a^3\dot{\phi}\dot{G}\delta_G 
+ 8\dot{a}^3\xi_\phi f_G\dot{G}\delta_G 
+ 8\dot{a}^3\xi_\phi \dot{\phi} f_G\delta_\phi = 0.
\label{eq:XL}
\end{align}

To calculate the values of an unknown functions, Eq.~(\ref{eq:XL}) is divided according to unique combinations of derivatives: $\dot{a}$, $\dot{\phi}$, $\dot{a}^3\dot{\phi}$, $\dot{a}^3\dot{G}$, $\dot{a}^2\dot{\phi}\dot{G}$, $\dot{a}\dot{\phi}$, $\dot{a}\dot{G}$, $\dot{\phi}\dot{G}$, $\dot{a}^2\dot{\phi}^2$, $\dot{a}^2\dot{G}^2$, $\dot{a}^4$. After simplifying and combining the equations, the following system is obtained
\begin{equation}
    \beta + 2a\beta_a = 0,
    \label{eq:cond1}
\end{equation}
\begin{equation}
    3\beta + 2a\delta_\phi = 0,
    \label{eq:cond2}
\end{equation}
\begin{equation}
    \left(3V + 3\xi f - 3\xi f_GG \right)\beta
    - a\xi f_{GG}G\gamma + \left(aV_\phi 
    + a\xi_\phi f - a\xi_\phi f_GG\right)\delta = 0,
    \label{eq:cond3}
\end{equation}
\begin{equation}
    \xi_\phi f_{GG}\gamma + \xi_{\phi\phi}f_G\delta 
    + 3\xi_\phi f_G\beta_a + \xi f_{GG}\gamma_\phi 
    + \xi_\phi f_G\delta_\phi = 0,
    \label{eq:cond4}
\end{equation}
\begin{equation}
    \xi f_{GGG}\gamma + \xi_{\phi}f_{GG}\delta 
    + 3\xi f_{GG}\beta_a + \xi f_{GG}\gamma_G = 0,
    \label{eq:cond5}
\end{equation}
\begin{equation}
    \beta_\phi = 0, \quad \beta_G = 0,
    \label{eq:cond6}
\end{equation}
\begin{equation}
    \delta_a = 0, \quad \delta_G = 0,
    \label{eq:cond7}
\end{equation}
\begin{equation}
    \gamma_a = 0.
    \label{eq:cond8}
\end{equation}
It follows from Eq.~(\ref{eq:cond1}) that the generator $\beta = \beta_0 a^{-1/2}$, where $\beta_0$ is an integration constant. However, there is a strong possibility that this form of $\beta$ does not satisfy the other equations. Substituting the expression obtained for $\beta$ into Eq.~(\ref{eq:cond2}) yields $\delta = \frac{3\beta\phi}{2a} + \delta_0$, where $\delta_0$ is an integration constant. Condition (\ref{eq:cond7}), however, imposes a restriction requiring that the generator $\delta$ must be independent of the scale factor $a(t)$. All consistency conditions are satisfied only if $\beta_0 = 0$, resulting in $\beta = 0$ and $\delta = \delta_0$. Substituting these values directly into Eq.~(\ref{eq:cond4}) and integrating with respect to the scalar field $\phi$ yields
\begin{equation}\label{4+}
    \xi f_{GG}\gamma = -\xi_\phi f_G \delta_0 - \phi_0,
\end{equation}
where $\phi_0$ is an integration constant. 

Since $\beta = 0$, the equation reduces to $V_\phi = 0$, implying a constant scalar potential $V(\phi) = V_0$. Substituting all these expressions for the generators and functions into Eq.~(\ref{eq:cond3}) yields the simplified form
\begin{equation}\label{3+}
    -\xi f_{GG}G\gamma + \xi_\phi f \delta_0 - \xi_\phi f_G G \delta_0 = 0.
\end{equation}

Substituting expression (\ref{4+}) into Eq.~(\ref{3+}) determines the form of the non-minimally coupled combination $\xi(\phi)f(G)$. As a result, the following set of solutions is obtained
\begin{align}
    &\beta = 0, \delta = \delta_0, V(\phi) = V_0, \nonumber \\
    &\xi(\phi)f(G) = -\frac{\phi_0}{\delta_0}G\phi + \phi_1,
    \label{NoetherSolution}
\end{align}
where $\phi_1$ is an integration constant. 
The solutions obtained show that, in order to recover the standard EsGB (\ref{standardEsGB}), it is sufficient to set $\phi_1 = 0$. In this case, $\xi(\phi) \propto \phi$ and $f(G) \propto G$. However, for a comprehensive analysis of the generalized EsGB model (\ref{actionmodel}), the integration constant is retained as $\phi_1 \neq 0$. The resulting values of the functions and generators make it possible to obtain a new symmetrized Lagrangian from (\ref{Lagrangian}).
\begin{equation}\label{Lagrangian*}
    L^* = -3 \dot{a}^2a + \frac{1}{2}a^3\dot{\phi}^2 - a^3V_0 -
    a^3 \phi_1 - 8\frac{\phi_0}{\delta_0}\dot{a}^3\dot{\phi} + a^3 L_m
\end{equation}

Since the new Lagrangian does not contain time derivatives higher than first order ones, the application of Noether symmetry methods leads to the elimination of ``ghosts", which is especially important in cosmological theories with higher-order invariants. The Euler-Lagrange equation and the ``zero" energy condition are used to analyse equations based on Lagrangian (\ref{Lagrangian*}).
\begin{align}
    &L^*_a - \left( L^*_{\dot{a}} \right)_t = 0 \rightarrow \nonumber 
    \\
    \rightarrow \;\;&3H^2 + 2\dot{H} + \frac{1}{2}\dot{\phi}^2 -V_0 - \phi_1 +
    16H\left(\dot{H}+H^2\right)\dot{\phi}\frac{\phi_0}{\delta_0} + 8H^2\ddot{\phi}\frac{\phi_0}{\delta_0} = 0
    \label{Friedmann1*} \\
    &L^*_\phi - \left( L^*_{\dot{\phi}} \right)_t = 0 \rightarrow \nonumber 
    \\
    \rightarrow \;\;&\ddot{\phi} + 3H \dot{\phi} -\frac{\phi_0}{\delta_0}G = 0
    \label{KleinGordon*} \\
    &L^*_{\dot{a}} \dot{a} + L^*_{\dot{G}} \dot{G} + 
    L^*_{\dot{\phi}} \dot{\phi} - L^* = 0 \rightarrow \nonumber 
    \\
    \rightarrow \;\;&-3H^2 + \frac{1}{2}\dot{\phi}^2 + V_0 + \phi_1 - 
    24H^3 \dot{\phi}\frac{\phi_0}{\delta_0} + \rho_m =0
    \label{Friedmann2*}
\end{align}

For the final analysis, it is necessary to compute the conserved quantity $Q$ and to verify its validity. To this end, the previous expressions for the generators (\ref{NoetherSolution}) and the Lagrangian (\ref{Lagrangian*}) are now
\begin{equation}
    Q = \beta L^*_{\dot{a}} + \gamma L^*_{\dot{G}} + 
    \delta L^*_{\dot{\phi}}  = \delta_0 \left( a^3 \dot{\phi} - 
    8 \dot{a}^3 \frac{\phi_0}{\delta_0}\right)
    \label{Q}
\end{equation}

The proof of the constancy of a conserved quantity (\ref{Q}) is based on the zero value of the first derivative with respect to cosmic time, namely
\begin{equation}
    \dot{Q} = a^3 \delta_0 \left( \ddot{\phi} + 3H\dot{\phi} -
    \frac{\phi_0}{\delta_0}G \right)
    \label{Qproof}
\end{equation}
The expression (\ref{Qproof}) matches the modified Klein-Gordon equation (\ref{KleinGordon*}), thereby confirming that the function $Q(t)=q_0=const$. This conservation law can be used to obtain the scalar field function form. The first derivative $\dot{\phi}$ of the scalar field is used, which is sufficient to substitute into the equations of motion (\ref{Friedmann1*}-\ref{Friedmann2*}).
\begin{equation}\label{dotphi}
    \dot{\phi} = 8H^3\frac{\phi_0}{\delta_0} + \frac{q_0}{a^3 \delta_0}
\end{equation}
To find solutions, Eq.~(\ref{Friedmann1*}) is equated to Eq.~(\ref{Friedmann2*}). This combination is acceptable, since both equations are independent and equal to zero, and there is no reduction in the constants $V_0$ and $\phi_1$. The matter energy density in these solutions has the form $\rho_m(t) = \rho_{m 0}a^{-3}$, where $\rho_{m0} = 3H_0^2 \Omega_{m0}$ is the matter energy density  at present, and $\Omega_{m0}$ is the matter density parameter. To proceed with further analysis of the obtained solutions and comparison of the results with the observational data, the transition to redshift terms $a(t) =(z+1)^{-1}$ is applied \cite{Bolotin2012ExpandingSpeedup}. 

The transition from time derivatives to derivatives with respect to the redshift is made according to the known rule $\frac{d}{dt} = -H(z)(1+z)\frac{d}{dz}$. After all these transformations and simplifications, the combined equation of motion takes the form
\begin{align}
    &2 (V_0 + \phi_1) + \rho_{m0}(1+z)^3 - \frac{2H}{\delta_0^2}
    \Big[ H \Big( 3 \delta_0^2 + 8\phi_0H \left( q_0 (1+z)^3 +20 \phi_0H^3\right)\Big)-
    \nonumber \\
    &-(1+z) \Big( \delta_0^2 + 8\phi_0H \left( q_0 (1+z)^3 +20 \phi_0H^3\right) H'\Big)
    \Big] = 0,
    \label{combinedEqMot}
\end{align}
where $H' = \frac{dH(z)}{dz}$. The solution to this equation is the Hubble parameter
\begin{equation}\label{HubbleSolution1}
    H(z)=\frac{1}{2 \cdot 5^{\frac{1}{3}}}
    \sqrt{
    \frac{5^{\frac{1}{3}} \delta_0^{\frac{2}{3}}
    \left(A + \sqrt{\frac{\delta_0^2}{40 \phi_0^2} + A^2}\right)^{\frac{1}{3}}}
    {\phi_0^{\frac{2}{3}}}
    -\frac{\delta_0^{\frac{4}{3}}}
    {2\phi_0^{\frac{4}{3}} 
    \left( A + \sqrt{\frac{\delta_0^2}{40 \phi_0^2} + A^2} \right)^{\frac{1}{3}}} 
    },
\end{equation}
where $A(z) = \frac{3h_0}{\delta_0^2}(1+z)^6 + \rho_{m0} (1+z)^3 + (V_0 + \phi_1)$, and $h_0$ is an integration constant. This function structurally resembles the square of the Hubble parameter in the standard $\Lambda$CDM model. In particular, one of its components, $\rho_{m0}(1+z)^3$, directly corresponds to the matter contribution in the Friedmann equation, where
\begin{equation}
    \rho_{m0}(1+z)^3 = 3H_0^2 \Omega_{m0}(1+z)^3.
\end{equation}
\begin{equation}\label{HubbleLCDM}
    H_{\Lambda \text{CDM}}(z)^2 = H_0^2 \left[ \Omega_{r0}(1+z)^4 + \Omega_{m0}(1+z)^3 + \Omega_{\Lambda 0}\right],
\end{equation}
with $\Omega_{r0}$, $\Omega_{m0}$ and $\Omega_{\Lambda 0}$ denoting the present time density parameters for radiation, matter, and dark energy, respectively. It can be seen that each component in $A(z)$ mirrors the redshift dependence of a specific physical component in $H_{\Lambda CDM}$. The term $(1+z)^3$ corresponds to matter and the constant terms $(V_0 + \phi_1)$ may be identified with an effective cosmological constant. 

The component $(1+z)^6$ is often obtained as a result of solutions with a predominance of scalar field kinetics \cite{Ferreira1998CosmologyField, Caprini2020CosmologicalWaves}. Components with this dependence on redshift are usually called ``stiff" matter. Even though the coefficients of the models differ, the structural similarity motivates making transformations of the form
\begin{align}
    \frac{3h_0}{\delta_0^2} (1+z)^6 &\rightarrow 3H_0^2 \Omega_{s0} (1+z)^6,\nonumber
    \\
    (V_0 + \phi_1) &\rightarrow 3H_0^2 \Omega_{\Lambda0},
    \label{transformations} 
\end{align}
where $\Omega_{s0}$ corresponds to the present time density parameter of a hypothetical stiff matter component. After these transformations, the final form of the Hubble parameter (\ref{HubbleSolution1}) reads
\begin{equation}
H(z) = \frac{1}{5^{1/3}}
    \sqrt{
    \frac{6 \cdot 5^{1/3} H_0^2  \Omega_{s0}^{1/3} \phi_0^{4/3} 
    \left( B + \sqrt{ B^2 + \frac{1}{1080 \phi_0^2 \Omega_{s0} H_0^6} } \right)^{2/3} - \phi_0^{2/3}}
    {24\ \phi_0^2\Omega_{s0}^{2/3} H_0^2 
    \left( B + \sqrt{B^2 + \frac{1}{1080 \phi_0^2 \Omega_{s0} H_0^6} } \right)^{1/3}}
    },
\label{HubbleSolution2}
\end{equation}
where $B(z) = \Omega_{\Lambda0} + \Omega_{m0} (1 + z)^3 + \Omega_{s0} (1 + z)^6$. In the next section, the solutions obtained are analyzed against the background of observational data and the transformations (\ref{transformations}) are justified.

\section{Observational Constraints and Model Testing}\label{sec4}

The resulting solution contains components similar to $\Lambda$CDM, but the modified form of the Hubble parameter (\ref{HubbleSolution2}) is different from the standard case (\ref{HubbleLCDM}). Therefore, it is necessary to prove the viability of the model and to compare it with the standard $\Lambda$CDM. The proof strategy is based on observational data from the late and early Universe.

To limit the model within the framework of the late expansion, three datasets are used, including data from Cosmic Chronometers (CC), Baryon Acoustic Oscillations (BAO), and data on Type Ia supernovae. This choice facilitates full-fledged analysis due to the fundamental independence and differences in the methods of obtaining observational data \cite{Adhikary2025DarkCosmology, Odintsov2025Einstein-Gauss-BonnetObservations}. Data from the Planck mission and the Atacama Cosmological Telescope (ACT) are used to constraint the model in the early stages. The analysis of the spectral index of scalar perturbations $n_S$ and the tensor-to-scalar ratio $r$ will be used. To substantiate the viability of the solutions obtained, a model stability analysis based on the speed of sound and an analysis of the equation of state parameter are used. This approach makes it possible to evaluate not only individual parameters of the model, but also the viability of the equations of motion.

The PolyChord sampler installed in the Cobaya \cite{Torrado2021Cobaya:Models, ASCL.netCosmology} package was used for Bayesian parameter analysis. PolyChord implements the nested sampling algorithm, designed to work efficiently in multidimensional parameter spaces and provides both the construction of a posterior distribution and an estimate of the Bayesian factor for comparing models \cite{Handley2015PolyChord:Cosmology, Handley2015PolyChord:Sampling}. The physical feasibility of solutions within a flat Universe is set using constraint $\Omega_m+\Omega_{\Lambda}+\Omega_s=1$.

\subsection{Late Universe Observational datasets}\label{subsec4.1}

\textit{Cosmic Chronometers} (CC). The dataset contains 32 independent measurements of the Hubble parameter $H(z)$ in the redshift range $0 < z  \lesssim 2$ \cite{Jalilvand2023Model-independentChronometers}. This method relies on estimating the differential age evolution of the Universe via the relation
\begin{equation}
    H(z) = -\frac{1}{1+z}\frac{dz}{dt}.
\end{equation}

This approach makes use of the relative age differences $\Delta t$ between galaxies with small redshift separation $\Delta z$ to infer $dz/dt$. The objects for measurements are early-type massive passive galaxies that do not form new stars, which makes it possible to accurately reconstruct their age from the spectrum. A major advantage of this dataset is its minimal dependence on cosmological assumptions, providing a model-independent probe of the expansion history at late times. For a given cosmological model, the theoretical prediction $H_{model}(z_i)$ is evaluated at the same redshift as the observations, and the log-likelihood is 
\begin{equation}\label{loglikeCC}
    \ln\mathcal{L}^{CC} = -\frac{1}{2} \chi^2= -\frac{1}{2} \sum_{i} \left( \frac{H_{model}(z_i) - H_i}{\sigma_{H_i}} \right)^2.
\end{equation}

\textit{Baryon Acoustic Oscillations} (BAO) represent a characteristic scale in the large-scale structure of the Universe, originating prior to the epoch of recombination as a result of sound wave propagation in the photon-baryon plasma. This scale corresponds to the sound horizon at recombination, $r_d \approx 147.4$ Mpc, and serves as a standard ruler for cosmological distance measurements.

Two main datasets are used. The first is the Sloan Digital Sky Survey (SDSS), which includes measurements from the Baryon Oscillation Spectroscopic Survey (BOSS) and the extended Baryon Oscillation Spectroscopic Survey (eBOSS) programs \cite{Alam2020TheObservatory}. This dataset provides 8 independent BAO measurements in the redshift range $0.15 < z < 2.33$. The second dataset is from the Dark Energy Spectroscopic Instrument (DESI), which offers 7 high-precision BAO measurements in the redshift range $0.295 < z < 2.33$ \cite{Adame2024DESIOscillations}. The Hubble parameter is reconstructed from the luminosity distance $d_L$
\begin{align}
    d_L(z) &= (1+z)\int_0^z \frac{c}{H(z')} \, dz'. \label{dL}\\
    \frac{D_M(z)}{r_d} = \frac{d_L(z)}{r_d (1+z)^2},\quad
    \frac{D_H(z)}{r_d} &= \frac{c}{r_d H(z)},\quad
    \frac{D_V(z)}{r_d} = \frac{1}{r_d} \left[ \frac{z\,d_L^2(z)}{(1+z)^2} \cdot \frac{c}{H(z)} \right]^{1/3},
\end{align}
where $D_H(z) = c/H(z)$ is the Hubble distance and $D_V(z)$ is the volume-averaged (isotropic) BAO distance. The log-likelihood functions used in the analysis are
\begin{align}
    \ln \mathcal{L}_{D_M} &= -\frac{1}{2} \sum_i 
    \left[ \frac{D_M(z_i) - D_M^{\text{obs}}(z_i)}{r_d\,\sigma_{D_M}(z_i)} \right]^2, \\
    \ln \mathcal{L}_{D_H} &= -\frac{1}{2} \sum_i
    \left[ \frac{\frac{c}{H(z_i)} - D_H^{\text{obs}}(z_i)}{r_d\,\sigma_{D_H}(z_i)} \right]^2, \\
    \ln \mathcal{L}_{D_V} &= -\frac{1}{2} 
    \sum_i \left[ \frac{D_V(z_i) - D_V^{\text{obs}}(z_i)}{r_d\,\sigma_{D_V}(z_i)} \right]^2, \\
    \ln \mathcal{L}^{\text{BAO}} &= 
    \ln \mathcal{L}_{D_M}^{\text{SDSS}} + \ln \mathcal{L}_{D_H}^{\text{SDSS}} + \ln \mathcal{L}_{D_V}^{\text{SDSS}} \nonumber \\
    &+ \ln \mathcal{L}_{D_M}^{\text{DESI}} + \ln \mathcal{L}_{D_H}^{\text{DESI}} + \ln \mathcal{L}_{D_V}^{\text{DESI}}.
    \label{loglikeBAO}
\end{align}

\textit{Pantheon+ with Supernovae and H0 for the Equation of State of dark energy} (Pantheon+ with SH0ES). The dataset contains data on 1701 light curves of 1550 unique, spectroscopically confirmed Type Ia supernovae (SNe Ia) in the redshift range $0 < z\leq 2.3$ \cite{Scolnic2022TheRelease}. The main feature of the dataset is the SH0ES calibration. It ensures consistency between the local measurements of the Hubble parameter and the cosmological parameters. The log-likelihood function based on the distance modulus $\mu(z)$ and the luminosity distance $d_L(z)$ (\ref{dL})
\begin{equation}\label{mu}
    \mu(z) = 5\; \text{log}_{10} \left( d_L(z) \right)+25.
\end{equation}

The calculated distance modulus (\ref{mu}) is used in the formula
\begin{equation}
    \text{ln} \mathcal{L}^{\text{Pantheon+ with SH0ES}} = -\frac{1}{2}\chi^2 
    = \left( \mu_{obs} - \mu_{model}\right)^T \mathbf{C}^{-1} \left( \mu_{obs} - \mu_{model} \right),
\end{equation}
where $\mathbf{C}$ is the complete covariance matrix incorporating both statistical and systematic uncertainties. For the stability of the calculations, the Cholesky decomposition $\mathbf{C} = \mathbf{L} \mathbf{L}^T$ with the lower triangular matrix $\mathbf{L}$ is used \cite{Lewis2013EfficientParameters}. This approach effectively computes $\chi^2$ without explicitly inverting the covariance matrix.
\begin{equation}
    \chi^2 = ||\mathbf{L}^{-1}\left( \mu_{obs} - \mu_{model}\right)||^2
\end{equation}

\subsection{Early Universe constraints}\label{subsec4.2}

An important feature of the EsGB models is the ability to meet the constraints of the early and late Universe at the same time. However, studying the slow-roll parameters of the early Universe in terms of redshift can lead to a noticeable loss of accuracy. To eliminate errors and align with observational data, a transition to e-folding terms is applied $a = \text{e}^N$, where $N$ is the number of e-foldings. The corresponding derivatives are calculated according to rule $\frac{d}{dt}=H(N)\frac{d}{dN}$ \cite{Navo2020StabilityStages}. Two main parameters are used to diagnose the model, which are typical for most models with a scalar field \cite{Hwang2004ClassicalAnalyses}.
\begin{equation}\label{slow-roll}
    \epsilon_1 = -\frac{\dot{H}}{H^2} \;,\;\; \epsilon_2 = - \frac{\ddot{\phi}}{H\dot{\phi}}.
\end{equation}

The slow-roll parameters constitute a key method for verifying the viability of the model, as they reveal the model's ability to support the concept of inflation. Meeting condition $\epsilon_1\ll 1$ proves the presence of inflation in the model, and meeting constraint $\epsilon_2\ll 1$ proves the sufficient duration of the inflationary period. Next, the slow-roll parameters (\ref{slow-roll}) are used to calculate the spectral index of scalar perturbations $n_S$ and the tensor-to-scalar ratio $r$, whose values can be compared with observational data \cite{Hwang2004ClassicalAnalyses}.
\begin{align}\label{spectralindeces}
    n_S-1 &= 4 \epsilon_1 - 2 \epsilon_2,\\
    r &= 4 \epsilon_1
\end{align}
The spectral index $n_S$ characterizes the dependence of primary scalar perturbations on the corresponding scale and it is an accurately measured parameter in the framework of observations of the cosmic microwave background (CMB). The ratio $r$ characterizes the ratio of the amplitudes of the tensor perturbations to the amplitude of scalar perturbations.

\textit{Planck 2018}. The work uses data from the Planck legacy archive \cite{PlanckArchive} and the chains obtained as part of the CamSpecHM + TTTEEE + lowl + lowE + BK15 + post + BAO + lensing configuration. It contains high-multipole CamSpec likelihoods in temperature and polarization (TT, TE, EE), the low--$\ell$ temperature and polarization likelihoods, BICEP2/Keck 2015 (BK15) constraints on primordial B-modes, baryon acoustic oscillations (BAO), and the reconstructed lensing spectrum \cite{Akrami2018PlanckInflation}. This combination provides the most stringent bounds on the scalar spectral index $n_s$ and the tensor-to-scalar ratio $r$, which are essential for testing inflationary models. 

\textit{Atacama Cosmology Telescope} (ACT). The DR6.02 2025 data release with measurements of temperature and polarization anisotropy \cite{Calabrese2025TheModels, LAMBDAMaps} are here used. ACT probes smaller angular scales than Planck, thereby offering an independent and complementary dataset that refines constraints on $n_s$ and $r$, and helps to test the robustness of inflationary predictions within the EsGB framework.

\subsection{Information Criteria}

Information criteria in the framework of statistical analysis are designed to compare models with different numbers of parameters and structures. For a full-fledged analysis, a set of three information criteria and an additional proof parameter $\log Z$ are used, each of which is able to identify the influence of additional parameters and the quality of the data approximation.

\textit{Akaike Information Criterion} (AIC). The Akaike criterion is based on evaluating the quality of a model through likelihood, adjusted for the number of parameters. It minimizes the expected Kullback–Leibler divergence between the true data distribution and the model \cite{Zhang2023InformationSelection, Dziak2019SensitivityCriteria} 

\begin{equation} 
\mathrm{AIC} = -2 \ln(L_{max}) + 2k,
\end{equation}
where $L_{max}$ is the maximum value of the likelihood function and $k$ is the number of model parameters. Lower values of the criterion indicate a more preferable model.

\textit{Bayesian Information Criterion} (BIC). The Schwartz criterion is a Bayesian analogue of the AIC. A special feature of the criterion is a more severe punishment for models with a large number of parameters. As in the previous case, lower values indicate a more preferred model \cite{Zhang2023InformationSelection, Dziak2019SensitivityCriteria}
\begin{equation}
    \mathrm{BIC} = -2 \ln(L_{max}) + k \ln(n).
\end{equation}

\textit{Deviance Information Criterion} (DIC). The deviance criterion is used primarily in a Bayesian context, especially when applying MCMC methods \cite{Liddle2007InformationSelection}. It measures the average fit and complexity of the model through deviations
\begin{equation}
    DIC = D(\bar{\theta}) + p_D, \;\;\;\; p_D = \overline{D(\theta)} - D(\bar{\theta}), \;\;\;\; D(\bar{\theta}) = -2\ln L(\bar{\theta}),
\end{equation}
where $D(\bar{\theta})$ is the deviation at the point of the average a posterior distribution of the parameters of $\bar{\theta}$, $p_D$ is the effective number of parameters and $\overline{D(\theta)}$ is the average deviation value according to the a posterior distribution. Lower values of the criterion indicate a more preferred model.

\textit{Logarithm of the evidence} ($\log Z$). Full Bayesian analysis uses the evidence Z, the logarithm of which plays a key role in calculating Bayesian factors
\begin{equation}
   \mathrm{BF}_{12} = \frac{p(y \mid M_1)}{p(y \mid M_2)} = e^{\log Z_1 - \log Z_2}. 
\end{equation}
The differences $\Delta\log Z$ serve as a Bayesian analogue of the information criteria \cite{Knuth2015DigitalSelection}. The higher $\log Z$, the better the model explains the data taking into account a priori distributions. The PolyChord sampler implements the nested sampling method, which directly evaluates the Bayesian evidence
\begin{equation}
    Z = \int L(\theta)\, \pi(\theta)\, d\theta, \quad \log Z \approx \sum_i\log\left(L_i\, w_i \right),
\end{equation}
where $L(\theta) = p(y \mid\theta)$, $\pi(\theta)$ is the a priori distribution and \(w_i\) is the step weight related to the volume of the remaining probability space. PolyChord iteratively updates the set of ``living points'' and calculates the logarithm of the evidence as the sum of the contributions in the likelihood space. PolyChord also evaluates uncertainty \(\sigma_{\log Z}\), which allows quantitative comparison of models through the difference \(\Delta\log Z\).

\section{Results and Discussion}\label{sec5}

In our analysis, prior constraints are imposed to define the admissible parameter space for the most probable values of the model parameters. Given the similarity of the  solution here considered to the $\Lambda$CDM model with its well-established cosmological bounds, the priors are specified as $50 < H_0 < 90$, $ 0.5 < \Omega_\Lambda < 0.9$ and $0.1 < \Omega_m < 0.5$. According to studies of stiff matter, the density parameter $\Omega_s$ has very small values, which makes its influence on the dynamics of the Universe significant in extremely early epochs, only. 

Since the effect on the later stages is insignificant, it is impossible to calculate the value of the parameter of stiff matter with high accuracy, within the framework of this study. Despite this fact, the presence of additional parameters is statistically important and it is used in the analysis of information criteria. For this reason, the limit is set as $0 < \Omega_s < 10^{-22}$ \cite{Gron2024CanTension, Dutta2010BigFluid}. Preliminary calculations indicate a weak effect of the $\phi_0$ parameter on the dynamics of the late expansion. The parameter boundaries are set as $0<\phi_0<1$.

\subsection{Parameter estimation and comparison}\label{subsec5.1}
\begin{table}[ht]
\centering
\caption{Posterior constraints (mean $\pm$ 1$\sigma$) on the cosmological parameters of the $\Lambda$CDM and EsGB models obtained from CC, BAO (SDSS+DESI), and Pantheon+ with SH0ES datasets.\label{CC+BAO+SnIa}}
\vspace{3mm}
{\begin{tabular}{@{}ccccccc@{}} \toprule
Dataset & Model & $H_0$ & $\Omega_{\Lambda}$ & $\Omega_m$ & $\Omega_s\cdot 10^{-24}$ & $\phi_0$ \\ \midrule
CC
& $\Lambda$CDM & 68.18$\pm$4.41 & 0.684$\pm$0.057 & 0.316$\pm$0.057 & --            & --              \\
& EsGB         & 68.26$\pm$4.35 & 0.681$\pm$0.057 & 0.319$\pm$0.057 & 50.6$\pm$28.3 & 0.505$\pm$0.289 \\
\midrule
BAO
& $\Lambda$CDM    & 68.34$\pm$1.61 & 0.694$\pm$0.033 & 0.306$\pm$0.033 & --            & --              \\
(SDSS+DESI)
& EsGB            & 68.31$\pm$1.55 & 0.694$\pm$0.032 & 0.306$\pm$0.032 & 51.1$\pm$28.6 & 0.502$\pm$0.288 \\
\midrule
Pantheon+
& $\Lambda$CDM & 72.84$\pm$0.50 & 0.638$\pm$0.036 & 0.362$\pm$0.036 & --            & --              \\
(with SH0ES)
& EsGB         & 72.85$\pm$0.53 & 0.639$\pm$0.039 & 0.361$\pm$0.039 & 51.7$\pm$28.7 & 0.548$\pm$0.256 \\
\bottomrule
\end{tabular}}
\end{table}

\begin{figure}[ht!]
\centerline{\includegraphics[width=0.75\textwidth]{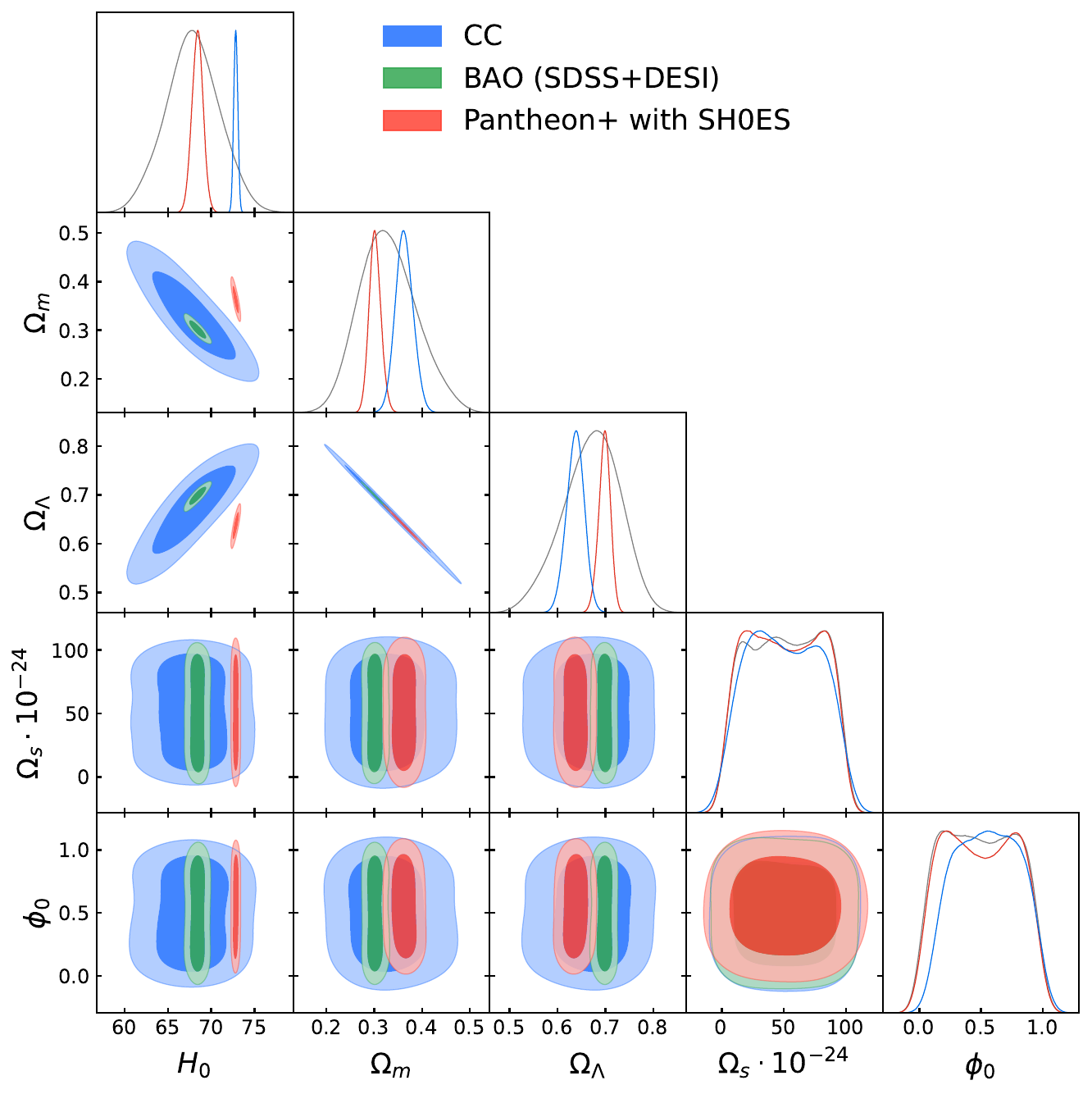}}
\caption{Posterior parameter distributions for the generalized EsGB derived from the CC (blue), BAO (SDSS+DESI) (green), and Pantheon+ with SH0ES (red) datasets, showing 1D marginalized posteriors (diagonal) and 1$\sigma$/2$\sigma$ confidence contours (off-diagonal)
}\label{Fig1}
\end{figure}

Table \ref{CC+BAO+SnIa} shows the results of the calculations of the parameters of the EsGB model and $\Lambda$CDM, based on the observational data. The results confirm the validity of transformations (\ref{transformations}), since the values of the parameters $\Omega_m$ and $\Omega_{\Lambda}$ are identical. The results of all datasets for both models indicate values of the Hubble constant $H_0 \approx 68-73$ $km\;s^{-1}Mpc^{-1}$, which correspond to the data obtained in other studies. The EsGB model did not have a noticeable effect on the dynamics of expansion in the late Universe relative to $\Lambda$CDM. The parameters $\Omega_s$ and $\phi_0$ also have a weak effect on the late dynamics, which is noticeable over wide confidence intervals. 

In both models, there are differences in the values of the Hubble constant $H_0$ when using different datasets, especially noticeable between Pantheon+ with SH0ES and other data. This factor is more clearly demonstrated in Fig.\ref{Fig1}. Despite the compliance of the parameters with the $\Lambda$CDM, the model contains additional parameters that may degrade the quality of the model. To this end, an analysis of information criteria was carried out to compare the models (Table \ref{InfoCriteria}).
\begin{table}[ht]
\centering
\caption{Information criteria (AIC, BIC, DIC) and log-evidence ($\log Z$) for the $\Lambda$CDM and EsGB models obtained from CC, BAO (SDSS+DESI), and Pantheon+ with SH0ES datasets. The differences $\Delta$ are defined as $\Delta = \Lambda$CDM $-$ EsGB.\label{InfoCriteria}}
\vspace{3mm}
{\begin{tabular}{@{}ccccccc@{}} \toprule
Dataset & Model & $\log L_{\max}$ & AIC & BIC & DIC & \quad $\log Z$ \\ \midrule
CC
& $\Lambda$CDM & $-48.28$ & $102.57$ & $106.97$ & $171.80$ & $-18.071 \pm 0.088$ \\
& EsGB         & $-43.62$ & $97.25$  & $104.58$ & $164.24$ & $-18.297 \pm 0.070$ \\
& $\Delta$     & $-4.22$  & $5.32$   & $2.39$   & $7.56$   & \quad $0.226 \pm 0.112$ \\
&              &          &          &          &          & Bayes factor: 1.254 \\
\midrule
BAO (SDSS+DESI)
& $\Lambda$CDM    & $-50.95$ & $107.89$ & $109.81$ & $172.77$ & $-27.521 \pm 0.135$ \\
& EsGB            & $-48.08$ & $106.16$ & $109.35$ & $160.45$ & $-27.624 \pm 0.104$ \\
& $\Delta$        & $-4.52$  & $1.74$   & $0.46$   & $12.32$  & \quad $0.104 \pm 0.170$ \\
&                 &          &          &          &          & Bayes factor: 1.109 \\
\midrule
Pantheon+ (with SH0ES)
& $\Lambda$CDM & $-66.25$ & $138.49$ & $154.81$ & $151.78$ & $-891.031 \pm 0.198$ \\
& EsGB         & $-61.73$ & $133.45$ & $160.65$ & $143.38$ & $-891.025 \pm 0.157$ \\
& $\Delta$     & $-4.36$  & $5.04$   & $-5.84$  & $8.40$   & $-0.006 \pm 0.253$ \\
&              &          &          &          &          & Bayes factor: 0.994 \\
\bottomrule
\end{tabular}}
\end{table}

Table \ref{InfoCriteria} contains information about the values of the information criteria (AIC, BIC and DIC) and the maximum logarithm of the likelihood function $\log L_{max}$, as well as the Bayesian evidence parameters obtained from the PolyChord sampler. The difference in statistical indicators between the two models is set as $\Delta = \Lambda$CDM$-$EsGB. The CC dataset fully supports the superiority of EsGB, indicated by the lower values of the likelihood function ($\Delta\log L_{\max} = -4.22$) and information criteria. The information criteria AIC ($\Delta$ = 5.32) and DIC ($\Delta$ = 7.56) indicate the average superiority of EsGB. A lower superiority of EsGB is observed in the BIC criterion ($\Delta$ = 2.39), despite the presence of additional parameters. The Bayesian factor ($\sim 1.25$), according to the Jeffreys scale \cite{Jeffreys1961TheoryProbability}, indicates a weak preference for the EsGB model.

Similar results are observed in BAO (SDSS+DESI) datasets, according to which the models do not have significant differences in AIC ($\Delta$ = 1.74) and BIC ($\Delta$ = 0.46), which is also evidenced by the Bayesian factor ($\sim 1.11$). However, DIC ($\Delta$=12.32) demonstrates a strong superiority of the EsGB model, which is consistent with the logarithm of the likelihood function ($\Delta\log L_{\max}= -4.52$).

Similar results were obtained using the Pantheon+ dataset with SH0ES calibration. The EsGB model is supported by the information criteria AIC ($\Delta$= 5.04) and DIC ($\Delta$=8.40), as evidenced by the likelihood function ($\Delta\log L_{\max}= -4.36$). However, BIC ($\Delta = - 5.84$) penalizes the model for more parameters, which is why it prefers $\Lambda$CDM. The Bayesian factor ($\Delta\sim 0.99$) does not give preference to any of the models. 

The analysis of information criteria and statistical characteristics have shown that the generalized EsGB model is not inferior to $\Lambda$CDM, while still having superiority according to some criteria. Our model exhibits competitiveness, even with two additional components. However, the analysis does not allow for identifying a clear favorite among the studied models.

The additional components of the model have little effect on the late dynamics, but their influence can be noticeable in the early stages. To this end, an analysis is carried out based on data from the Planck mission and ACT measurements; a graphical interpretation  is shown in Fig. \ref{EsGB_Planck_Act}. 
\begin{figure}[ht!]
\centerline{\includegraphics[width=0.6\textwidth]{Fig2.pdf}}
\vspace*{8pt}
\caption{Constraints on the tensor-to-scalar ratio $r$ and scalar spectral index $n_s$ from Planck 2018 (green) and ACT DR6.02 (blue) data, compared with the predictions of the EsGB model (red). The shaded regions correspond to the $68\%$ and $95\%$ confidence levels}\label{EsGB_Planck_Act}
\end{figure}

The model corresponds to the observational data only in the confidence interval 2$\sigma$. The values of the scalar spectral index $n_s=0.975^{+0.007}_{-0.004}$ correspond better to the ACT data, while there is a good correspondence to the Planck data, but in terms of the tensor-to-scalar ratio ($r=0.05^{+0.008}_{-0.015}$). 

In summary, the EsGB model remains consistent with current cosmological observations and shows moderate advantages in specific datasets. Although the improvements are not yet statistically decisive, this framework provides a promising extension capable of connecting late-time acceleration with early-Universe phenomena.

\subsection{Equation of State and Stability Analysis}\label{subsec5.2}

In the previous section, an analysis was performed using observational data, in which the Hubble parameter function (\ref{HubbleSolution2}) was tested. Despite the high degree of compliance with the observational data, the analysis did not fully take into account the behavior of the entire model, namely, the dynamics of the total density $\rho (z)$ and pressure $p(z)$ of the model based on the equations of motion
\begin{equation}\label{pressure}
    -(3H^2 + 2\dot{H})= p =  \frac{1}{2}\dot{\phi}^2 -V_0 - \phi_1 +
    16H\left(\dot{H}+H^2\right)\dot{\phi}\frac{\phi_0}{\delta_0} + 8H^2\ddot{\phi}\frac{\phi_0}{\delta_0}, 
\end{equation}
\begin{equation}\label{density}
    3H^2 = \rho = \frac{1}{2}\dot{\phi}^2 + V_0 + \phi_1 - 
    24H^3 \dot{\phi}\frac{\phi_0}{\delta_0} + \rho_m.
\end{equation}

Since the most probable values of all the parameters are known, substituting them into the equations of motion (\ref{pressure}, \ref{density}) will help analyze the expansion dynamics, taking into account all the components of the model. To evaluate possible states, the equation of state parameter (EoS) is used, calculated using $\omega=\frac{p}{\rho}$ formula \cite{Melia2014TheState}. The transition from decelerated to accelerated expansion occurs at $\omega=-\frac{1}{3}$. To check the consistency of the transition, the deceleration parameter $q(z)$ is used, which depends only on the Hubble parameter and redshift \cite{Bolotin2012ExpandingSpeedup, Chaudhary2024AddressingParametrization, Capozziello2019ExtendedCosmography}, namely
\begin{equation}
    q(z) = \frac{H'(1+z)}{H}-1.
\end{equation}

\begin{figure}[ht!]
\centerline{\includegraphics[width=0.6\textwidth]{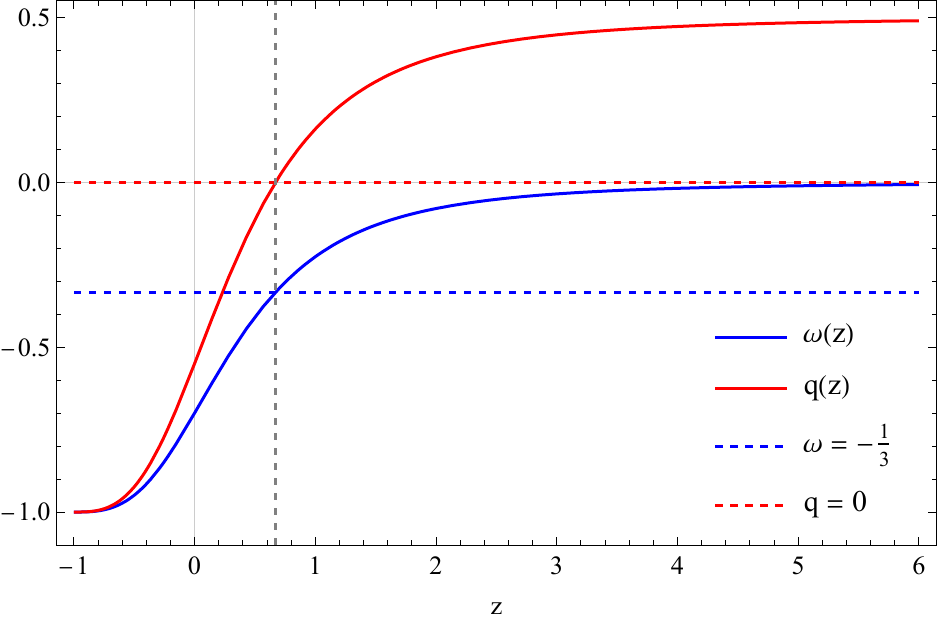}}
\caption{The equation of state parameter $\omega(z)$ (blue) and the deceleration parameter $q(z)$ (red) of the model. The corresponding dashed lines indicate the parameter values at the transition from decelerated to accelerated expansion}\label{EsGB_EoS_deceleration}
\end{figure}

The deceleration parameter indicates the presence of a transition at $q(z)=0$ \cite{Capozziello2008CosmographyGravity}. The graphical interpretation obtained based on the values of the model parameters is shown in Fig. \ref{EsGB_EoS_deceleration}.

From the model we obtain a transition from decelerated to accelerated expansion at $z \approx 0.66$. Moreover, there is no contradiction between the deceleration parameter and the EoS parameter, which indicates the internal consistency of the model.

\begin{figure}[ht!]
\centerline{\includegraphics[width=0.6\textwidth]{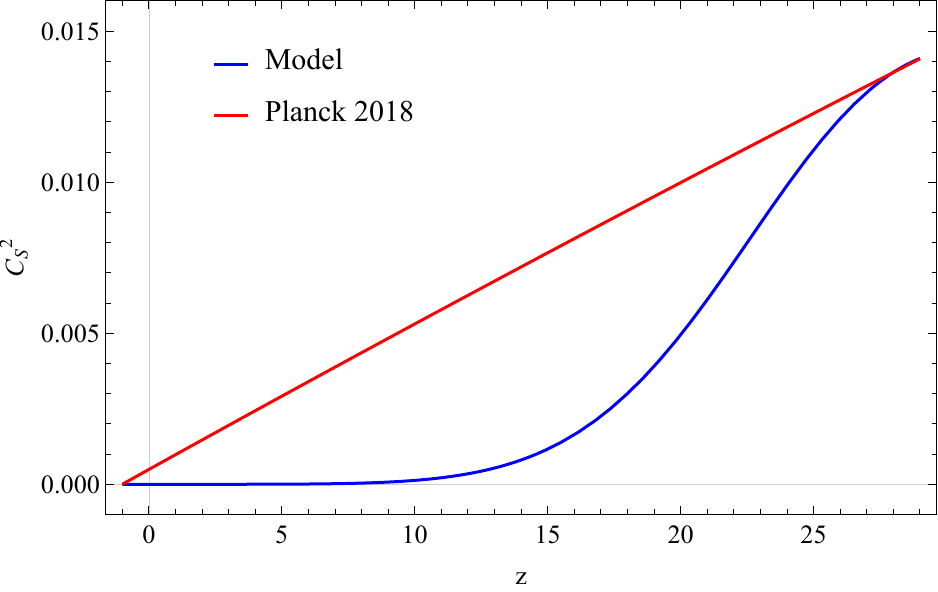}}
\caption{The square of the speed of sound $C_S^2(z)$ as a condition for the stability of the model (blue) in comparison with the results of observations Planck 2018 (red)}
\label{EsGB_Stab}
\end{figure}

Fig. \ref{EsGB_EoS_deceleration} demonstrates the existence of an era of matter dominance  ($\omega=0$), which smoothly transitions into the era of  dark energy dominance ($\omega=-1$), while not switching to the phantom energy mode ($\omega<-1$). To prove the stability of the model, the speed of sound $C_S$ is used \cite{Shaily2025BouncingFRLm-gravity}, which is calculated by means of the formula $C_S^2 = \frac{\dot{p}}{\dot{\rho}}$. Stable models must meet $C_S^2>0$ condition, which is demonstrated in the EsGB model in Fig. \ref{EsGB_Stab}. In addition to the positive values of the speed of sound, it is important to check the correspondence of the values with the observational data. The formula below is used to calculate the model-independent sound speed \cite{Choueiri2008AcousticUniverse}
\begin{equation}
    C_S^2 = \frac{1}{3(1+R(z))}, \;\;\;\; R(z) = \frac{\rho_{b0}(1+z)^3}{\rho_{\gamma 0}(1+z)^4},
\end{equation}
where $\rho_{b0}$ is the current value of the density of baryon matter and $\rho_{\gamma 0}$ is the current value of the radiation density. According to the Planck 2018 data, $R(z) \approx 680 (1+z)^{-1}$

The model is stable up to large redshifts ($z\approx30$),  corresponding to the values of the speed of sound measured during the Planck mission. However, the values vary at lower redshifts. The model shows a faster decrease in the speed of sound, reaching near-zero values in the late Universe.

\section{Conclusion}\label{sec6}

A new analytic solution for the generalized EsGB model has been obtained. The use of Noether symmetry methods provided a very convenient way to determine the functional forms of all the components of the action and moreover, to obtain the conservation law, which led to a self-consistent form of the Hubble parameter. As a result, a model naturally arises that is quite close to $\Lambda$CDM, but exhibits additional components that modify the dynamics in the early stages of the Universe evolution. 

A closer analysis of the  solution  has shown that the introduction of Noether symmetry constraints eliminates derivatives in the Lagrangian above the second order, which leads to the disappearance of Ostrogradsky's ghosts, a very remarkable property. Moreover, the resulting dynamics maintain stability, as proven by the positive sound speed values. It also turns out that the model does not switch to the phantom mode, although it maintains the presence of a transition from decelerated to accelerated expansion. This is confirmed by the consistency of the deceleration parameter and of the EoS parameter. The transition moment perfectly matches with the one obtained from observational data, corresponding to a redshift of $z\approx0.66$.

Late-stage viability testing has been performed based on recent CC, BAO (SDSS, DESI) and Pantheon+ with SH0ES calibration data. The results of the statistical analysis showed a high degree of compliance of the model with the observational data. Moreover, most of the information criteria hint towards a clear, although weak, preference for our model  in comparison with $\Lambda$CDM. 

An additional analysis of the parameters of the early Universe also complies with the data from Planck 2018 and ACT, which definitely proves the  ability of our model to accurately describe both the early and late stages of the Universe evolution. To finish, our results confirm that Noether symmetry methods can serve as an effective tool in constructing stable theories of modified gravity that combine mathematical elegance and physical feasibility.

\vskip 0.8cm
\centerline{\bf Acknowledgments}
\vskip 0.2cm
\noindent
This research was funded by the Science Committee of the Ministry of Science and Higher Education of the Republic of Kazakhstan (Grant No. AP26194585). It has also been partially supported by the program Unidad de Excelencia María de Maeztu CEX2020-001058-M, and by the Catalan Government, AGAUR project 2021-SGR-00171.

\printbibliography

\end{document}